\documentclass[cits]{PoS}
\usepackage{amsthm,amsmath,exscale,epsfig,subfigure}
\usepackage[applemac]{inputenc}
\usepackage{placeins}

\title{Scaling study for 2 HEX smeared fermions: hadron and quark masses}

\ShortTitle{Scaling study for 2 HEX smeared fermions}

\author{\speaker{Thorsten Kurth}$^a$, Stephan Durr$^{ab}$, Zoltan Fodor$^{abc}$, Christian Hoelbling$^a$, 
Sandor D.~Katz$^{ac}$, Stefan Krieg$^{ab}$, Laurent Lellouch$^d$, Thomas Lippert$^{ab}$, Kalman K.~Szabo$^a$, Gregory Vulvert$^d$\\
        \llap{$^a$} Bergische Universität Wuppertal, Gaussstr. 20, D-42119 Wuppertal, Germany\\
        \llap{$^b$} Jülich Supercomputing Centre, Forschungszentrum Jülich, D-52425 Jülich, Germany\\
        \llap{$^c$} Institute for Theoretical Physics, Eötvös University, H-1117 Budapest, Hungary\\
        \llap{$^d$} Centre de Physique Théorique\footnote{CPT is research unit UMR 6207 of the CNRS and of the universities Aix-Marseille I, Aix-Marseille II and Sud Toulon-Var, and is affiliated with the FRUMAM.}, Case 907, Campus de Luminy, F-13288 Marseille, France\\
        E-mail: \email{thorsten.kurth@uni-wuppertal.de}}
\author{Budapest-Marseille-Wuppertal Collaboration}

\abstract{The goal of this study is to investigate the scaling behaviour of our 2 HEX action. For this purpose, we compute the $N_f=3$ spectrum and compare the results to our 6 EXP action. We find a large scaling window up to  $\sim 0.15\,\mathrm{fm}$ along with small scaling corrections at the $2\%$-level and full compatibility with our previous study. As a second important observable to be tested for scaling, we chose the non-perturbatively renormalized quenched strange quark mass.  Here we find a fairly flat scaling with a broad scaling range up to $\simeq 0.15\,\mathrm{fm}$ and perfect agreement with the literature.}

\FullConference{The XXVIII International Symposium on Lattice Field Theory, Lattice2010\\
		June 14-19, 2010\\
		Villasimius, Italy}

\begin{document}

\section{Definition of the 2 HEX action}
For the gauge part in the $N_f=3$ simulations, we use a tree-level improved Symanzik gauge action
\begin{equation}\label{gauge_action}
S_\mathrm{g}=\beta\left[\frac{c_0}{3}\sum\limits_{\mathrm{plaq}}\mathrm{Re}\mathrm{Tr}(1-U_{\mathrm{plaq}})+\frac{c_1}{3}\sum\limits_{\mathrm{rect}}\mathrm{Re}\mathrm{Tr}(1-U_{\mathrm{rect}})\right],
\end{equation}
where $c_1=-1/12$ and $c_0=1-8c_1=5/3$ and the standard Wilson plaquette action in case of the quenched study. For the fermionic part, we used a tree-level improved clover-action
\begin{equation}\label{fermion_action}
S_\mathrm{f}=S_{\mathrm{W}}-\frac{c_{\mathrm{SW}}}{2}\sum\limits_x\sum\limits_{\mu<\nu}(\bar{\psi}\sigma_{\mu\nu}F_{\mu\nu}[V]\psi)(x),
\end{equation}
with $c_{\mathrm{SW}}=1$ and $S_\mathrm{W}$ the standard Wilson action, coupled to smeared links $V$. These smeared links are constructed by combining the HYP setup \cite{Hasenfratz:2001hp} with the analytic stout (EXP) recipe \cite{Morningstar:2003gk}:
\begin{eqnarray}\label{hex-recipe}
\Gamma^{(1)}_{\mu:\nu\rho}(\mathbf{x})&=&\sum\limits_{\pm \sigma\neq (\mu,\nu,\rho)}U_\sigma(\mathbf{x})U_\mu(\mathbf{x}+\hat{\sigma})U^\dagger_\sigma(\mathbf{x}+\hat{\mu})\nonumber\\
V^{(1)}_{\mu:\nu\rho}(\mathbf{x})&=&\exp\left(\frac{\alpha_3}{2}\,\mathrm{P}_{\mathrm{TA}}\left\{\Gamma^{(1)}_{\mu:\nu\rho}(\mathbf{x}) U^\dagger_\mu(\mathbf{x})\right\}\right)U_\mu(\mathbf{x})\nonumber\\
\Gamma^{(2)}_{\mu:\nu}(\mathbf{x})&=&\sum\limits_{\pm \sigma\neq (\mu,\nu)}V^{(1)}_{\sigma:\mu\nu}(\mathbf{x})V^{(1)}_{\mu:\nu\sigma}(\mathbf{x}+\hat{\sigma})V^{(1)\,\dagger}_{\sigma:\mu\nu}(\mathbf{x}+\hat{\mu})\nonumber\\
V^{(2)}_{\mu:\nu}(\mathbf{x})&=&\exp\left(\frac{\alpha_2}{4}\,\mathrm{P}_{\mathrm{TA}}\left\{\Gamma^{(2)}_{\mu,\nu}(\mathbf{x}) U^\dagger_\mu(\mathbf{x})\right\}\right)U_\mu(\mathbf{x})\nonumber\\
\Gamma^{(3)}_\mu(\mathbf{x})&=&\sum\limits_{\pm \nu\neq \mu}V^{(2)}_{\nu:\mu}(\mathbf{x})V^{(2)}_{\mu:\nu}(\mathbf{x}+\hat{\nu})V^{(2)\,\dagger}_{\nu:\mu}(\mathbf{x}+\hat{\mu})\nonumber\\
V_\mu(\mathbf{x})&=&\exp\left(\frac{\alpha_1}{6}\,\mathrm{P}_{\mathrm{TA}}\left\{\Gamma^{(3)}_\mu(\mathbf{x}) U^\dagger_\mu(\mathbf{x})\right\}\right)U_\mu(\mathbf{x}).
\end{eqnarray}
This smearing was introduced in \cite{Capitani:2006ni}. Here we chose the weights $(\alpha_1,\alpha_2,\alpha_3)=(0.95,0.76,0.38)$.
It is straightforward to show that this smearing is analytic with respect to the thin links $U_\mu$ and hence can be used within an HMC for dynamical simulations. We applied the recipe (\ref{hex-recipe}) twice to define our 2 HEX action. 
Generically, such smearing improves the chirality of the underlying Dirac operator and drives renormalization constants closer to their tree-level values \cite{Capitani:2006ni}. Hence we expect a scaling behaviour which is close to that of a non-perturbatively $\mathcal{O}(a)$ improved action, although our action is formally $\mathcal{O}(\alpha\,a)$ improved.

\section{Scaling of $N_f=3$ hadron masses}
\begin{figure}[ht!]
\subfigure[$\mathcal{O}(\alpha a)$-scaling]{
\includegraphics[scale=0.6]{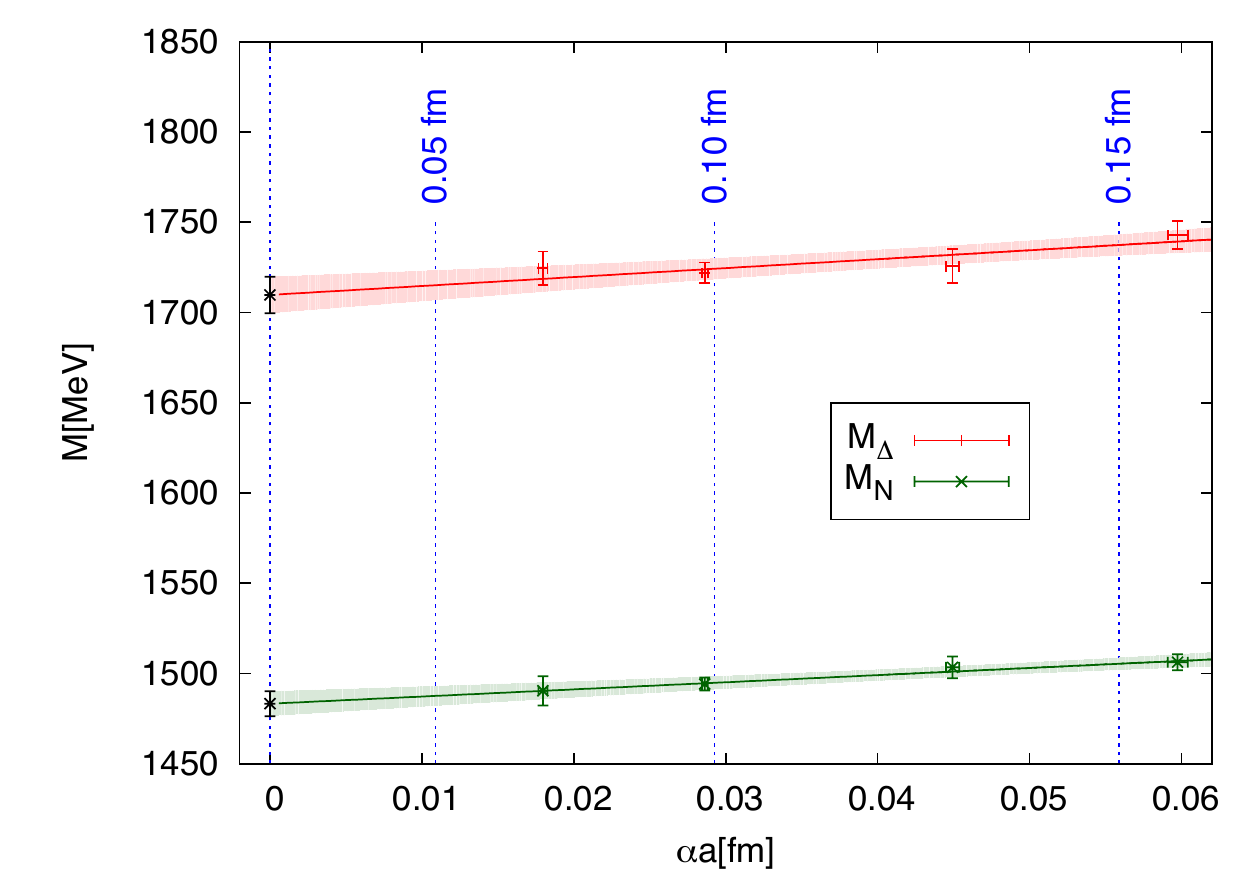}
}
\subfigure[$\mathcal{O}(a^2)$-scaling]{
\includegraphics[scale=0.6]{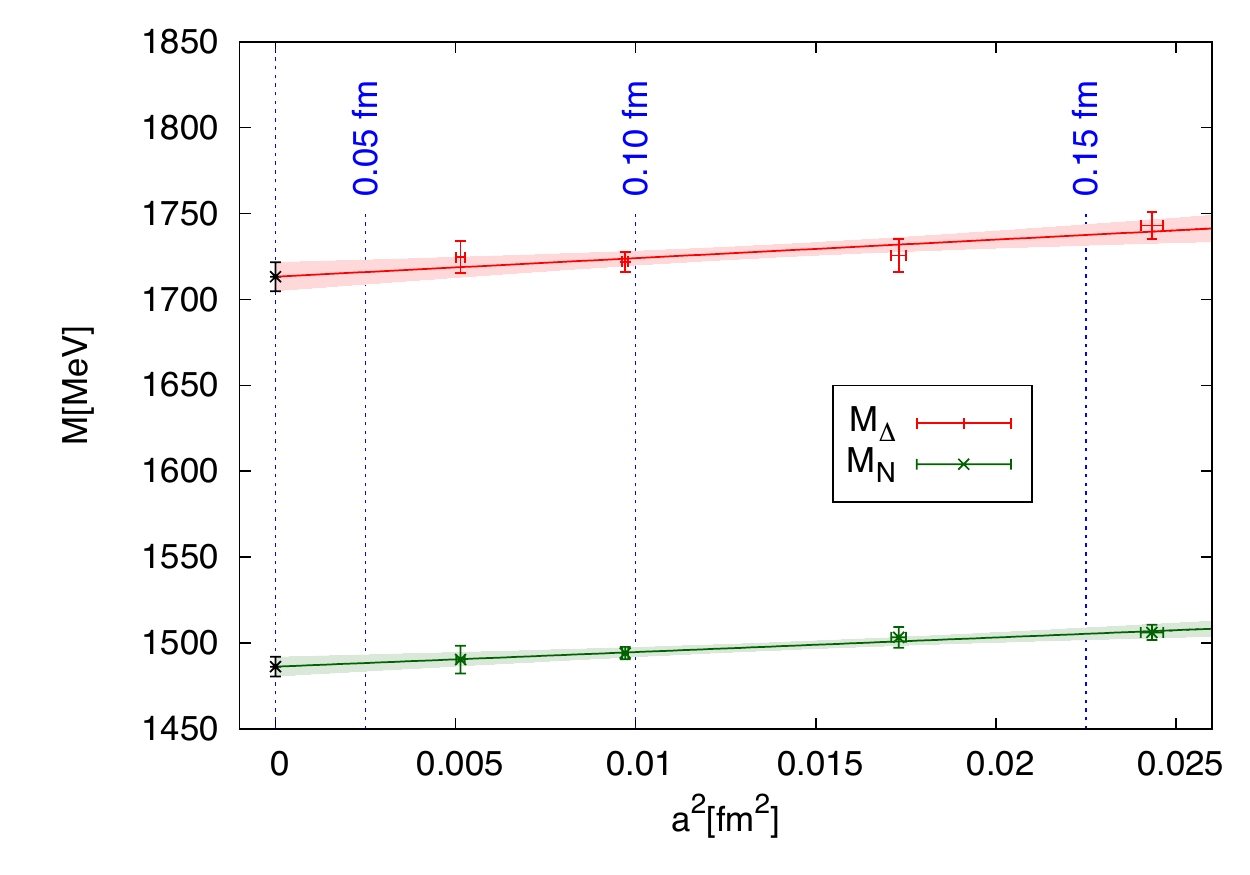}
}
\caption{\label{hadron-scaling}Continuum extrapolation of the octet and decouplet masses assuming different scaling.}
\end{figure}

As a first test for this action, we computed the $N_f=3$ hadron spectrum to be able to compare the results to our former 6 EXP action \cite{Durr:2008rw}. We took four values for $\beta$ so that the cutoff varied between $a\approx 0.06\,\mathrm{fm}$ and $0.2\,\mathrm{fm}$. At each beta we simulated at least four different masses to be able to perform a safe interpolation to our reference point $M_\pi/M_\rho\doteq \big[2(M_K^{\mathrm{phys}})^2-(M_\pi^{\mathrm{phys}})^2\big]^{1/2}/M_\phi^{\mathrm{phys}}\approx 0.67$. 
The hadron masses themselves were obtained by applying correlated $\cosh$ or $\sinh$-fits to the correlators, where we avoided fitting excited state contributions by inspecting the effective mass plateau and chose the fit ranges accordingly. The PCAC mass was computed by fitting the plateau of $\langle \partial_0A_0\bar{P}\rangle/\langle P\bar{P}\rangle$. Then we extrapolated the octet and decouplet masses to the continuum assuming either $\mathcal{O}(\alpha a)$ or $\mathcal{O}(a^2)$ scaling, as displayed in Fig. \ref{hadron-scaling}. We see flat scaling with a maximum of $\sim 2\%$ scaling corrections for the delta mass around $a\simeq 0.16\,\mathrm{fm}$. The scaling window is broad and extends at least up to this lattice spacing. The fit qualities favor the $\mathcal{O}(a^2)$-scaling but in order to compute a reliable systematical error, the $\mathcal{O}(\alpha a)$-scaling assumption should also be taken into account.

\section{Scaling of quenched quark masses}
\begin{figure}[ht!]
\includegraphics[scale=1.1]{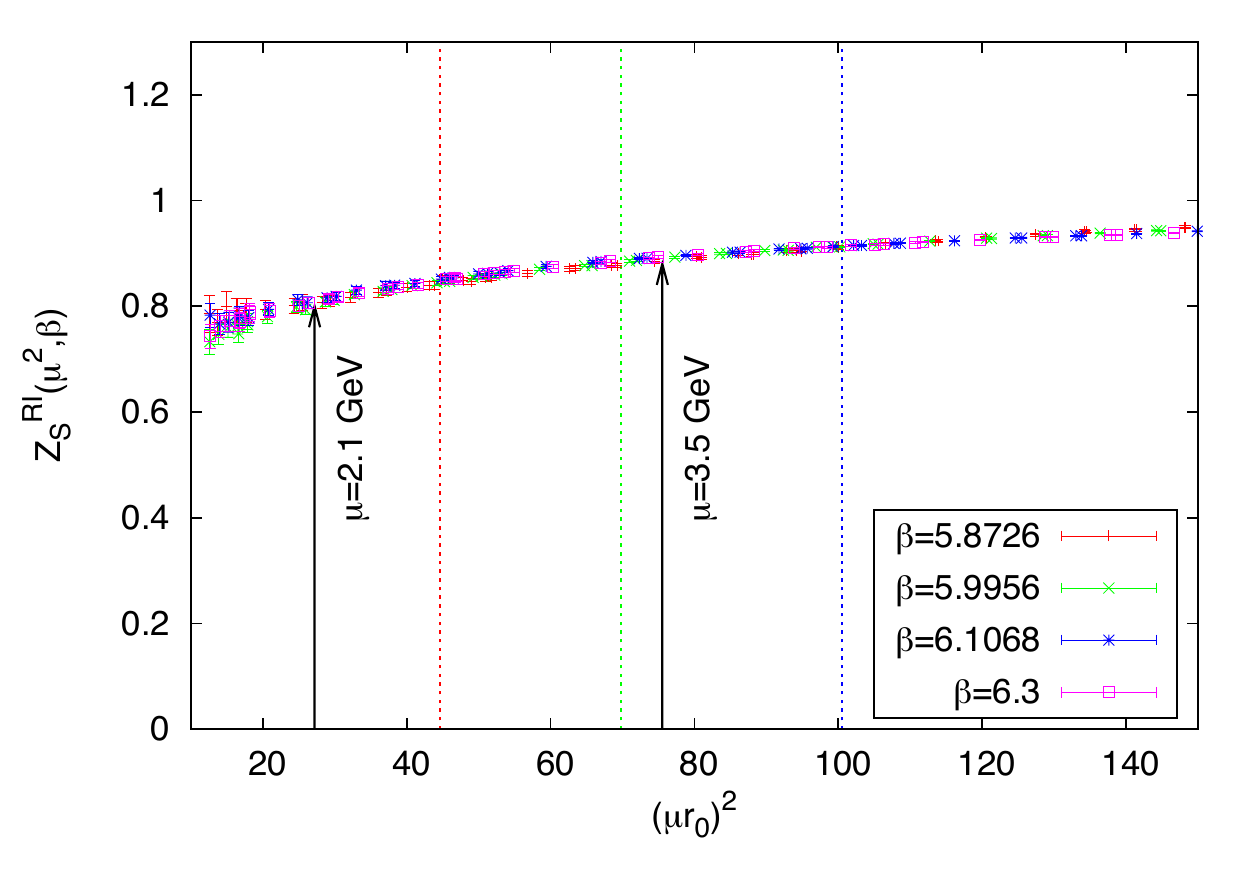}
\caption{\label{matched_zs}Multiplicatively matched $Z_S$. The colored dashed vertical bars denote the different cutoff $p=\pi/(2a)$ for the three coarsest lattice spacings. For $\beta=6.3$, $(\mu r_0)^2\approx 179$ is off the scale.}
\end{figure}

\begin{figure}[ht!]
\includegraphics[scale=1.1]{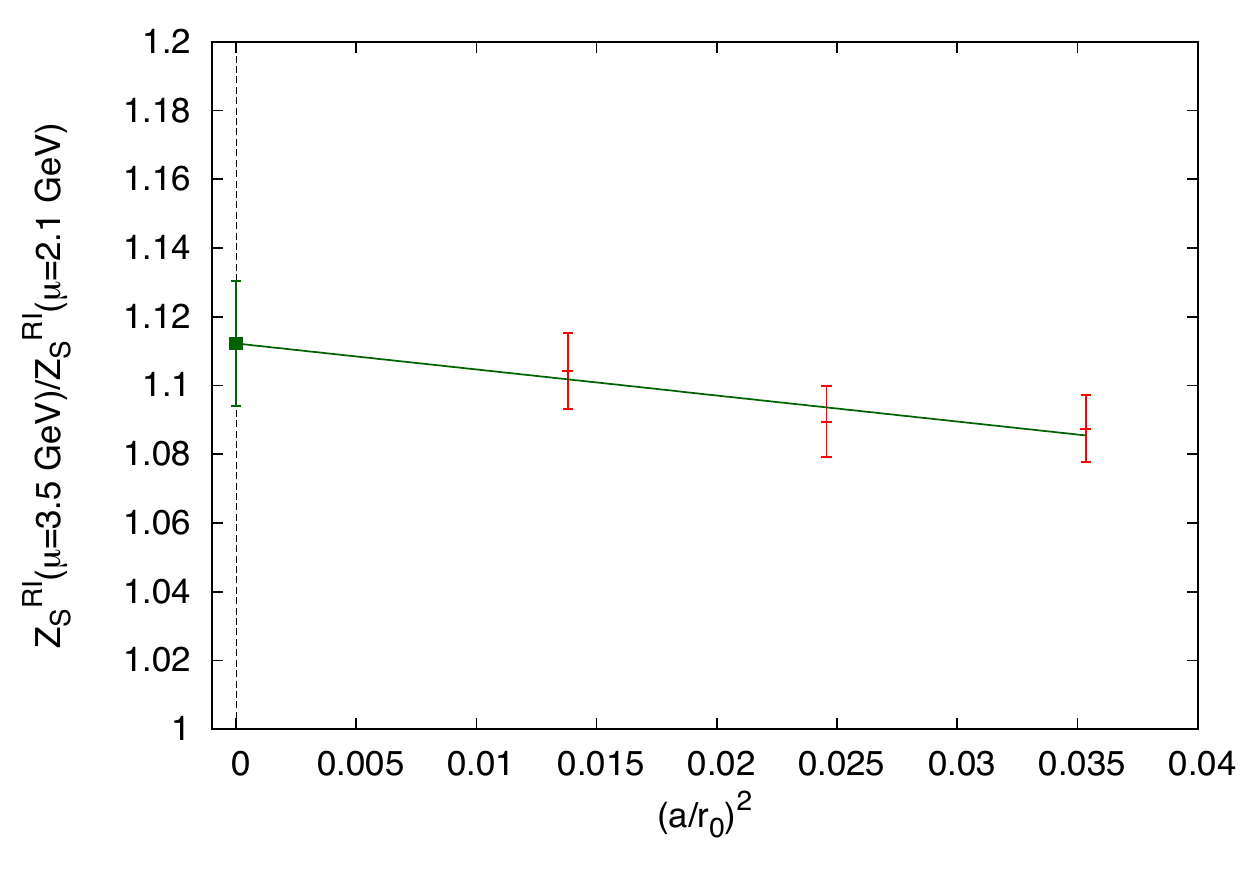}
\caption{\label{zs_ratio_extrapol_asq}Continuum limit of $Z_S(3.5\,\mathrm{GeV},a)/Z_S(2.1\,\mathrm{GeV},a)$.}
\end{figure}

\begin{figure}[ht!]
\subfigure[$\mathcal{O}(\alpha a)$-scaling]{
\includegraphics[scale=0.99]{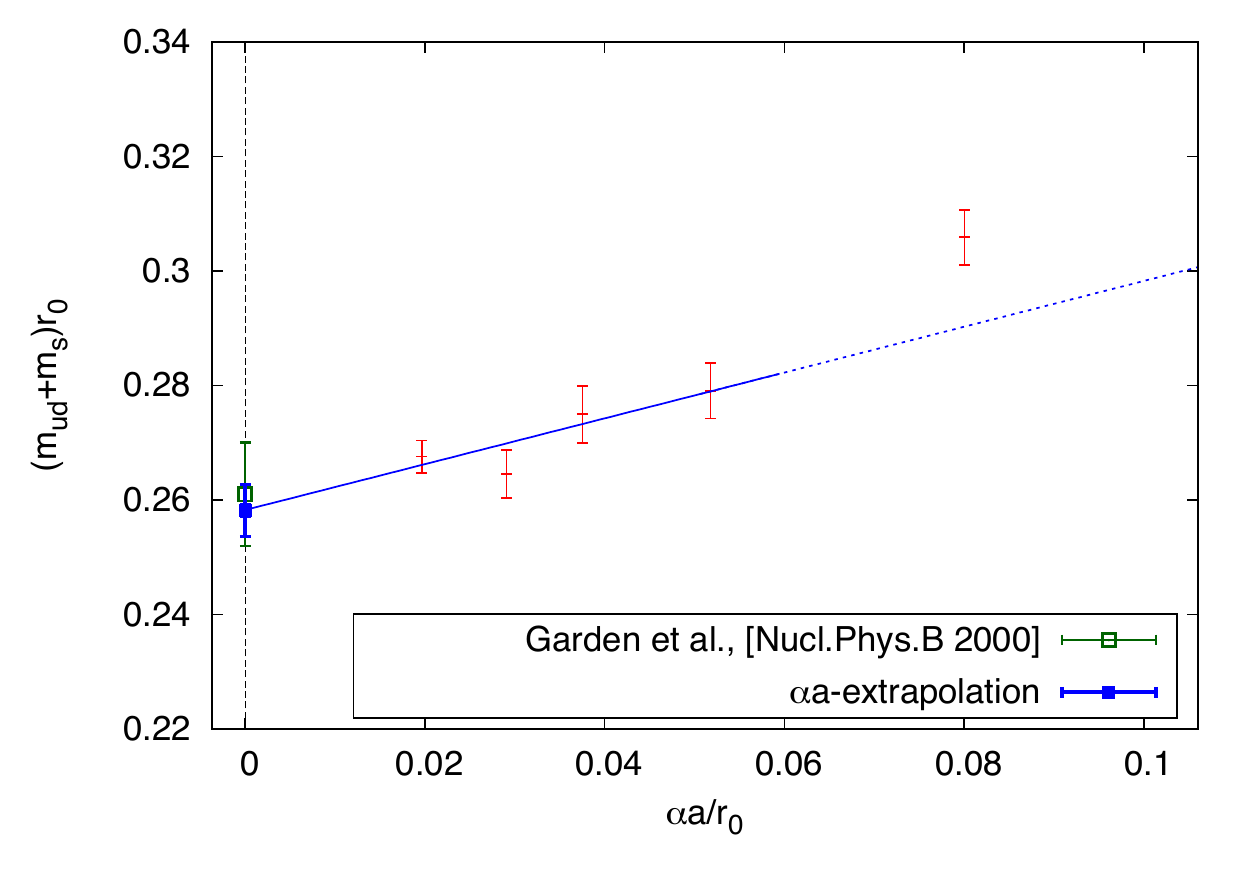}
}
\subfigure[$\mathcal{O}(a^2)$-scaling]{
\includegraphics[scale=0.99]{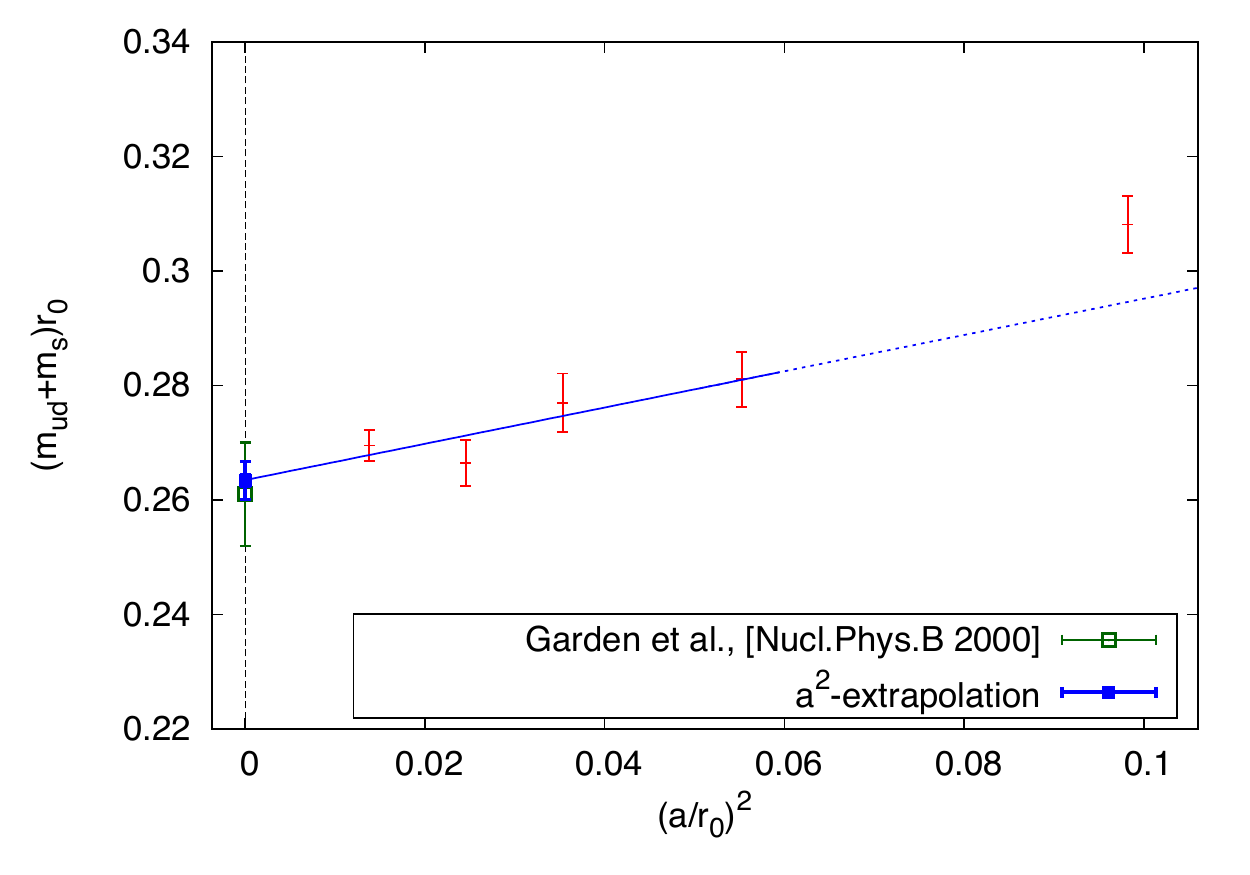}
}
\caption{\label{quark-scaling}Continuum extrapolation of the quenched $(m_s+m_{ud})^{\mathrm{RI}}$ at $2\,\mathrm{GeV}$ in units of $r_0$.}
\end{figure}

Another important observable to study is the strange quark mass. The renormalization factors have been computed by using the non-perturbative RI method \cite{Martinelli:1994ty}. The renormalization conditions are defined as follows:
\begin{equation}\label{renormcond}
Z_O(\mu a,g(a))\,Z^{-1}_Q(\mu a,g(a))\,\Gamma_O(pa)|_{p^2=\mu^2}=1,
\end{equation}
where $\Gamma_O(pa)=\mathrm{Tr}[(S^{-1}G_OS^{-1})(pa)\,P_O]$ is the renormalized vertex function belonging to operator $O$ with the unamputated Greens function $G_O$ and a projector $P_O$, which projects to the quantum numbers of $O$. In order to remove an $\mathcal{O}(a)$ contact term we applied the trace subtraction procedure described in \cite{Capitani:2000xi,Martinelli:2001ak,Maillart:2008pv}, therefore replacing all quark propagators $S$ by $\bar{S}=S-\mathrm{Tr}_DS/4$. The quark mass renormalization constant is $1/Z_S^{\mathrm{RI}}$, but the renormalization condition (\ref{renormcond}) only gives ratios $(Z_O/Z_Q)^{\mathrm{RI}}$.
In order to avoid discrete derivatives as they would appear in the calculation of the wave-function renormalization $Z_Q$ itself within the RI formalism, $Z_Q$ was computed indirectly by using the VWI. To be more precise, consider the ratio
\begin{equation}
\zeta(t)=\frac{\sum_x \langle P(T/2) V_4(x,t) \bar{P}(0)\rangle}{\langle P(T/2)\bar{P}(0)\rangle},
\end{equation}
with $P(x,t)=\sum_x \bar{\psi}_1\gamma_5\psi_2(x,t)$ and $V_4(x,t)=\sum_x \bar{\psi}_2\gamma_4\psi_2(x,t)$. The tree-level on-shell-improved vector renormalization constant can then be obtained by computing \cite{Martinelli:1993dq,Gockeler:1999cy}
\begin{equation}\label{zv-vwi}
Z_V(1+am^W)=[\zeta(t_0>T/2)-\zeta(t_0-T/2)]^{-1},
\end{equation}
where $m^W=m^{\mathrm{bare}}-m^{\mathrm{crit}}$. Hence the wave-function renormalization can be obtained via $Z_Q^{\mathrm{RI}}=(Z_Q/Z_V)^{\mathrm{RI}}\,Z_V$. We found this method to be more precise than changing to the RI$^\prime$ scheme or using the conserved vector current within the RI formalism.\\
Our data are reasonably consistent with perturbation theory for $p\geq 3\,\mathrm{GeV}$. We found that the influence of the cutoff on the RI data is small for momenta up to $p<\pi/(2a)$, i.e. the scale $p\approx 3\,\mathrm{GeV}$ was not safely reachable at the coarsest lattice as can be seen from Fig. \ref{matched_zs}. Hence we computed the ratio
\begin{equation}\label{ratiofunc}
R(\mu',\mu'')=\lim\limits_{a\rightarrow 0}\frac{Z_S(\mu',a)}{Z_S(\mu'',a)},
\end{equation}
for $\mu'=3.5\,\mathrm{GeV}$ and $\mu''=2.1\,\mathrm{GeV}$ only on the three finest lattices. The continuum limit was taken again assuming $\mathcal{O}(\alpha a)$ or $\mathcal{O}(a^2)$ scaling and is relatively flat (c.f. Fig. \ref{zs_ratio_extrapol_asq}), as suggested by Fig. \ref{matched_zs}. Using these relations, we defined the quenched quark mass via
\begin{equation}
m^{\mathrm{VWI}}(\mu')=(1-am^W/2)m^W/[R(\mu',\mu'')Z_S(\mu'',a)].
\end{equation}
The continuum extrapolations are displayed in Fig. \ref{quark-scaling}. As before, we considered both versions of scaling and see that both extrapolations are compatible and fairly flat, with at most $7\%$ scaling corrections up to $\sim 0.12\,\mathrm{fm}$. When considering the full dataset, the $\mathcal{O}(a^2)$ extrapolation is slightly favored. Our combined result is
\begin{equation}
[(m_s+m_{ud})r_0]^{\overline{\mathrm{MS}}}(2\,\mathrm{GeV})=0.2609(39)(28),
\end{equation}
where the first error is statistical and the second error systematical, not including the uncertainty due to quenching. This is result is compatible with \cite{Garden:1999fg} and in good agreement with other results from the literature. Assuming $m_s/m_{ud}\approx 27.5$ as suggested by recent unquenched lattice calculations and $r_0=0.49$ this translates into
\begin{equation}
m_s^{\overline{\mathrm{MS}}}(2\,\mathrm{GeV})=101.4(1.5)(1.1).
\end{equation}
Note, that for this value there is an unknown systematic error due to the above assumptions for $m_s/m_{ud}$ and $r_0$ and the systematic error analysis described below does not include these sources of uncertainties.\\
The statistical errors have been computed by using 2000 moving-block-bootstrap samples with atomic blocking because all configurations are well decorrelated. For handling the systematic errors, we followed the procedure described in \cite{Durr:2008zz}.
Thus we performed the analysis using three different fit ranges to extract the masses from the correlators, two different scaling assumptions for the continuum limit and three scales $\mu=3.0$, $3.5$ and $4.0\,\mathrm{GeV}$ at which we matched non-perturbative results onto perturbation theory. This yields $3\cdot 2\cdot 3=18$ different continuum limits which were all weighted by their quality of fits. The mean gives the overall estimate and the variance our systematic error.

\FloatBarrier

\section{Summary}
We performed hadron and quark masses scaling studies for our new 2 HEX action. We found large scaling windows and mild deviations from the continuum limit in both cases. Moreover, this action significantly reduces the condition number of the Wilson-Dirac operator for small quark masses. We conclude that this action should be well suited for large unquenched phenomenological studies.

\section{Acknowledgments}
This work was partly supported by the SFB TR55, by EU grant MRTN-CT-2006-035482 (FLAVIAnet) and by CNRS grants GDR n\textsuperscript{\tiny 0} 2921 and PICS n\textsuperscript{\tiny 0} 4707 as well as FO/502. All computations have been carried out on JUGENE and JUROPA at FZ Jülich and on clusters at the University of Wuppertal. We also like to thank Szabolcs Borsanyi for his technical help.


\begin{thebibliography}{99}

\bibitem{Hasenfratz:2001hp}
  A.~Hasenfratz and F.~Knechtli,
  Phys.\ Rev.\  D {\bf 64} (2001) 034504
  [arXiv:hep-lat/0103029].
  
\bibitem{Morningstar:2003gk}
  C.~Morningstar and M.~J.~Peardon,
  Phys.\ Rev.\  D {\bf 69} (2004) 054501
  [arXiv:hep-lat/0311018].
  
  \bibitem{Capitani:2006ni}
  S.~Capitani, S.~Durr, C.~Hoelbling,
  JHEP {\bf 0611 } (2006)  028.
  [hep-lat/0607006].
  
  \bibitem{Durr:2008rw}
  S.~Durr {\it et al.},
  Phys.\ Rev.\ D {\bf 79 } (2009)  014501.
  [arXiv:0802.2706 [hep-lat]].
  
\bibitem{Martinelli:1994ty}
  G.~Martinelli {\it et al.},
  Nucl.\ Phys.\ B {\bf 445 } (1995)  81-108.
  [hep-lat/9411010].

\bibitem{Capitani:2000xi}
  S.~Capitani, M.~Gockeler, R.~Horsley, H.~Perlt, P.~E.~L.~Rakow, G.~Schierholz and A.~Schiller,
  Nucl.\ Phys.\ B {\bf 593}, 183 (2001)
  [arXiv:hep-lat/0007004].

\bibitem{Martinelli:2001ak}
  G.~Martinelli {\it et al.},
  Nucl.\ Phys.\ B {\bf 611} (2001)  311-337.
  [hep-lat/0106003].
  
\bibitem{Maillart:2008pv}
  V.~Maillart and F.~Niedermayer,
  arXiv:0807.0030 [hep-lat].
 
  
\bibitem{Martinelli:1993dq}
  G.~Martinelli, S.~Petrarca, C.~T.~Sachrajda and A.~Vladikas,
  Phys.\ Lett.\ B {\bf 311} (1993) 241
  [Erratum-ibid.\ B {\bf 317} (1993) 660].

\bibitem{Gockeler:1999cy}
  M.~Gockeler {\it et al.},
  Phys.\ Rev.\ D {\bf 62} (2000) 054504
  [arXiv:hep-lat/9908005].
 
\bibitem{Garden:1999fg}
  J.~Garden, J.~Heitger, R.~Sommer and H.~Wittig  [ALPHA and UKQCD
                  Collaboration],
  Nucl.\ Phys.\ B {\bf 571} (2000) 237
  [arXiv:hep-lat/9906013].
 
\bibitem{Durr:2008zz}
  S.~Durr {\it et al.},
  Science {\bf 322}, 1224 (2008)
  [arXiv:0906.3599 [hep-lat]].
 
\end{thebibliography}
\end{document}